\documentclass{PoS}

\usepackage{amsmath}
\usepackage{amssymb}
\usepackage{amstext}
\usepackage{amsfonts}
\usepackage{subfigure}
\usepackage{multirow}

\newcommand{\ba}{\begin{eqnarray}}
\newcommand{\ea}{\end{eqnarray}}
\newcommand{\be}{\begin{equation}}
\newcommand{\ee}{\end{equation}}
\newcommand{\bd}{\begin{displaymath}}
\newcommand{\ed}{\end{displaymath}}
\newcommand{\een}{\nonumber\end{equation}}
\newcommand{\bea}{\begin{eqnarray}}
\newcommand{\eean}{\nonumber\end{eqnarray}}
\newcommand{\eea}{\end{eqnarray}}

\def\eq#1{Eq.~(\ref{#1})}

\def\fig#1{Fig.~\ref{#1}}

\newcommand{\gap}{\hspace{10pt}}

\newcommand{\mev}{\mathrm{MeV}}

\newcommand{\fm}{\mathrm{fm}}

\newcommand{\mps}{m_{\rm{PS}}}

\newcommand{\beq}{\begin{equation}}   
\newcommand{\eeq}{\end{equation}}   
\newcommand{\beqn}{\begin{eqnarray}}  
\newcommand{\eeqn}{\end{eqnarray}}

\def\mcC{{\mathcal C}}
\def\mcD{{\mathcal D}}
\def\mcJ{{\mathcal J}}

\def\mcO{{\mathcal O}}

\def\la{\langle}
\def\ra{\rangle}
\def\psibar{\overline{\psi}}
\def\chibar{\overline{\chi}}

\newcommand{\old}[1]{}

\title{\vspace*{-2cm}{\large \normalfont \hfill DESY 11-213, SFB/CPP-11-64} \\\vspace*{1cm}
Nucleon scalar matrix elements with $N_f=2+1+1$ twisted mass fermions\vspace*{0.3cm}
\begin{center}
 \includegraphics[scale=0.2]{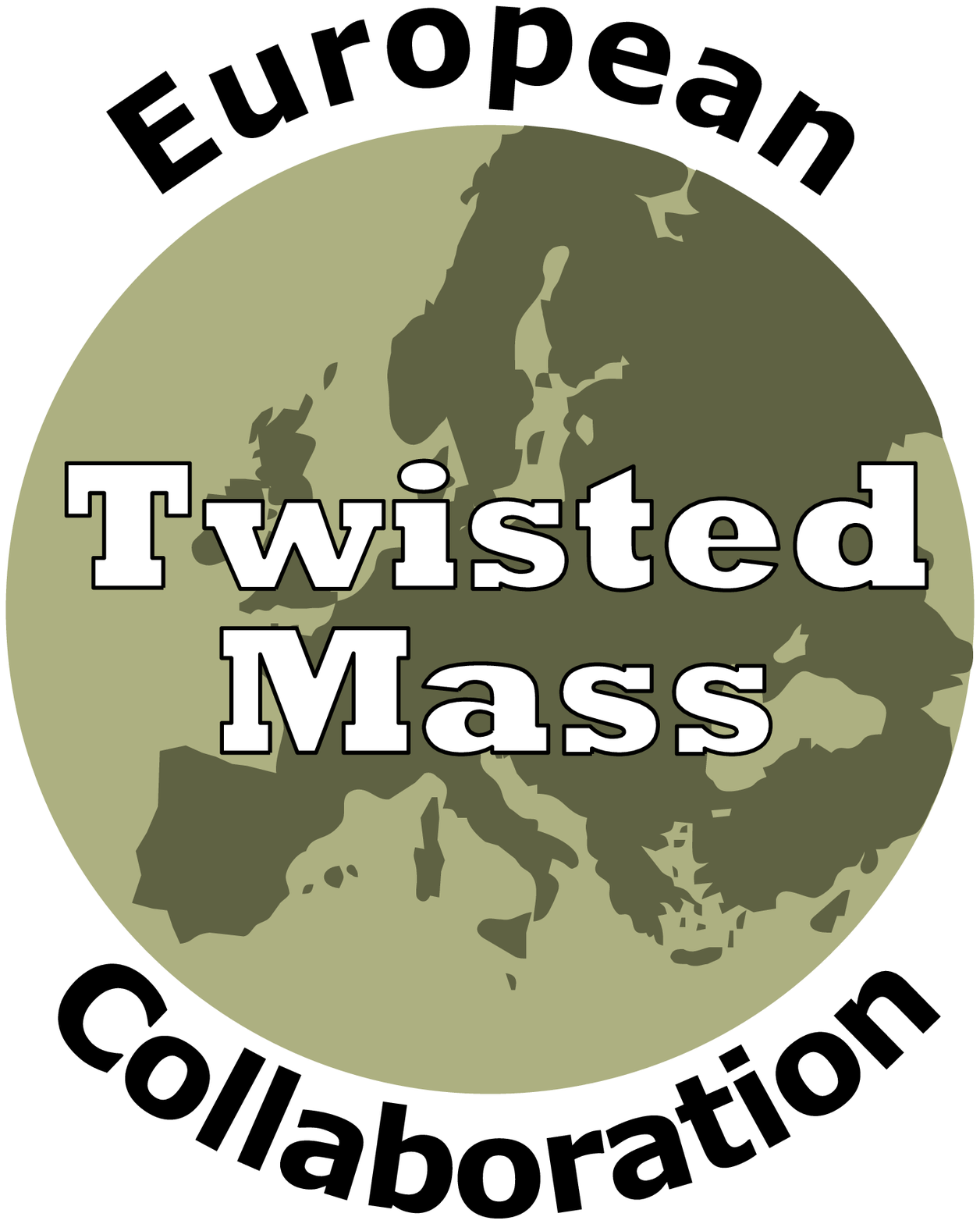}
\end{center}
}

\ShortTitle{Nucleon scalar matrix elements with  twisted mass fermions}

\author{Simon Dinter$\,^a$,
\speaker{Vincent Drach}$\,^a$,
Karl Jansen$\,^a$ \\
\llap{$^a$}{NIC, DESY Zeuthen, Platanenallee 6, D-15738 Zeuthen, Germany\\}

E-mail: \email{vincent.drach@desy.de}

}

\abstract{We investigate scalar matrix elements of the nucleon using $N_f=2+1+1$ 
flavors of maximally twisted mass fermions at a fixed value of the lattice spacing 
of $a\approx 0.078{\rm fm}$.
We compute disconnected contributions
to the relevant three-point functions using an efficient noise reduction technique. 
Using these methods together with an only 
multiplicative renormalization applicable for twisted mass fermions, allows us to obtain 
accurate results in the light and strange sector. }

\FullConference{ The XXIX International Symposium on Lattice Field Theory - Lattice 2011\\
July 10-16, 2011\\
Squaw Valley, Lake Tahoe, California}

\begin{document}

\section{Introduction}

The various evidences of the existence of dark matter have led to the development of 
experiments dedicated to detect dark matter directly. The detection relies on the 
measurements of the recoil of atoms hit by a dark matter candidate. One popular class of 
dark matter models involve an interaction between a WIMP and a Nucleon mediated by a Higgs exchange.  
Therefore, the scalar content of the nucleon is a fundamental ingredient in the WIMP-Nucleon 
cross section. In this way, the uncertainties of the scalar content translates directly  
into the accuracy of the constraints on beyond the standard model physics. 
Since the coupling of the Higgs to quarks is proportional to the quark masses, 
it is important to know how large scalar matrix elements of the nucleon are, 
in particular for the strange and charm quarks.  

One common way to write the parameters entering the relevant cross section are the 
so-called sigma-terms of the nucleon:
\be\label{eq:sigma_terms}
\sigma_{\pi N} \equiv m \la N|  \bar{u} u+  \bar{d} d |N \ra \gap\textmd{and}\gap 
\sigma_0  \equiv  m\la N|  \bar{u} u+ \bar{d} d  - 2 \bar{s}s|N \ra\; ,
\ee
where $m$ denotes the light quark mass. Quantifying the scalar strangeness content of the nucleon 
a parameter $y_N$ is introduced,

\be\label{eq:y_para}
y_N \equiv  \frac{ 2\la N| \bar{s} s |N \ra }{ \la N| \bar{u} u+ \bar{d} d |N \ra},
\ee
which can be also related to the sigma terms of the nucleon in eq.~(\ref{eq:sigma_terms}).

The direct computation of the above matrix elements is known to be challenging on the lattice for 
several reasons. First, it involves the computation of "singlet"  or "disconnected'' diagrams 
that are very noisy on the lattice. Second, discretisations that break chiral symmetry generally 
suffer from a mixing under renormalization between the light and strange sector, 
which is difficult to treat in a fully non-perturbative way.  
However, twisted mass fermions offer two advantages here: they provide both an efficient variance 
noise reduction for disconnected diagrams and a convenient way to avoid the 
chirally violating contributions that are responsible for the mixing under renormalization.

\section{Lattice Setup}

In our simulations we use the mass-degenerate twisted mass action in the light sector and 
the mass non-degenerate twisted mass action in the strange and charm sector. 
The quark masses of the heavy quark doublet have been tuned such that the Kaon and D-mesons 
masses assume approximately their physical value. The reader interested in more details about 
aspects of this setup is referred to \cite{Baron:2010bv,Baron:2010th}. 
The twisted mass action in the light sector reads  

\bea
S[\chi,\chibar,U] &=&  \sum_{x} \chibar_q(x)D_{\rm tm}[U]  \chi_q(x)  =  \sum_{x} \Bigg\{ \chibar_q(x)(\frac{1}{2\kappa}+  i  a\mu_q \gamma_5 \tau^3 )\chi_q(x) \\
&-& \frac{1}{2} \chibar_q(x) \sum_{\mu=0}^3  \Big[ U_{\mu}(x) (r+\gamma_{\mu}) \chi_q(x + a\hat{\mu}) + U^{\dagger}_{\mu}(x-a\hat{\mu}) (r -\gamma_{\mu}) \chi_q(x- a\hat{\mu})\Big] \Bigg\} . \nonumber
\label{eq:twisted_action_nf2}
\eea
where the hopping parameter $\kappa=(2am_0 +8r)^{-1}$ is  defined in terms of 
$am_0$, the bare Wilson mass, $r$ is the Wilson parameter and 
$\mu_q$ is the bare twisted mass parameter.  The Wilson parameter is fixed to $|r|=1$ is all our simulations.  
When  $\kappa$ is tuned to its critical value a situation called 
maximal twist is achieved which guarantees 
$O(a)$ improvement of physical observables. 

%
For further needs we also introduce the operators $D_{q,\pm}$ denoting the 
upper and lower components of the Wilson 
twisted mass operator in flavour space (also referred to as the Osterwalder-Seiler Dirac operator):
\be
D_{q,\pm}[U] = {\bf{tr}}~\frac{1\pm\tau_3}{2}D_{\rm tm}[U] ,
\label{eq:Dpm}
\ee
where ${\bf{tr}}$ denotes the trace in flavour space.
$D_{q,\pm}[U]$ then corresponds to 1-flavour twisted mass operators with
Wilson parameter $r=\pm 1$.

When we discuss the 2-point and 3-point functions necessary for this work, 
we will use the so-called physical basis of quark fields denoted as $\psi_q$. 
This field basis is related to the twisted quark field basis, $\chi_q$ by 
the following field rotation
\be\label{eq:rotation_phys}
\psi_q \equiv e^{i\frac{\omega_l}{2} \gamma_5 \tau^3} \chi_q \gap\textmd{and}\gap  \psibar_q \equiv \chibar_q  e^{i\frac{\omega_l}{2} \gamma_5 \tau^3},
\ee
where the twist angle $\omega_l=\pi/2$ at maximal twist. 

In order to compute matrix elements involving strange quarks, we will work within a 
mixed action setup. For reasons that will become clear later, we choose to introduce 
in the valence sector an additional doublet of degenerate  twisted mass quark of mass $\mu_q$ . 
The mass $\mu_q$ can then be tuned to reproduce the Kaon and D-mesons mass in the unitary setup. 
Preliminary estimates of the matching masses gives  $a\mu_s=0.0185$ in the strange 
sector and $a\mu_c=0.2514 $ in the charm sector and we will approximately use these values for $\mu_q$ further on. 
In this contribution, we work at a fixed lattice spacing corresponding to 
$a \approx 0.078~\fm$ with $\mps L \ge  4$ and pion masses ranging approximately from $300$ to $500~\mev$.

\section{Matrix elements}

In the following, $\psi_q$ with index $l,s,d$ will denote the quark fields of the 
light ($l$), strange ($s$) or charm ($c$) quarks in the physical basis. 
In order to be self-contained, we describe in this section the 
relevant correlation functions used in this work. 
The nucleon two-point function reads :

\be\label{eq:C2pts}
C^{\pm }_{N,\rm 2pts}(t -  t_{\rm src},\vec{x}_{\rm src }) =   \sum_{\vec{x}}  {\bf{tr}}~
\Gamma^{\pm}  \la \mcJ_{N}(x) \overline{\mcJ_{N}}(x_{\rm src})  \ra ,
\ee
where $x_{\rm src}\equiv (t_{\rm src},\vec{x}_{\rm src})$ is the source 
point and the subscript $N$ refers to the proton or to the neutron states for 
which the interpolating fields are given by:
\bd
\mcJ_p = \epsilon^{abc} \left( u^{a,T} \mcC\gamma_5 d^b \right) u^c \gap\textmd{and}\gap   \mcJ_n = \epsilon^{abc} \left( d^{a,T} \mcC\gamma_5 u^b \right) d^c.
\ed
The projectors used are $\Gamma^{\pm} = \frac{1\pm\gamma_0}{2}$, and $\mcC$ is the 
charge conjugation matrix. 
 
The nucleon three-point functions is 
\be\label{eq:C3pts}
C^{\pm,O_q}_{N,\rm 3pts}(t_{s},\Delta t_{\rm{op}},\vec{x}_{\rm src }) =     
\sum_{\vec{x} ,\vec{x}_{\rm op}}   {\bf{tr}}~ 
\Gamma^{\pm}\la \mcJ_{N}(x) O_q(x_{\rm op}) \overline{\mcJ_{N}}(x_{\rm src})  \ra , 
\ee
where  $O_q$ is an operator having scalar quantum numbers, e.g. $O_q=\bar{q}q$,  
$\Delta t_{\rm op} =t_{\rm{op}} - t_{\rm{src}} $ is the time of insertion 
of the operator, and $t_s=t-t_{\rm src}$ gives the so-called source-sink 
separation. Note that in the twisted basis the scalar operators read
\be
\widetilde{O}_q=i\chibar_q\gamma_5 \tau^3 \chi_q,  \gap\textmd{where}\gap q=l,s,c~,
\ee
and are hence given by the pseudo scalar density. 
The flavour structure of the operators in the twisted basis will be crucial in the following.

\begin{figure}[htb]
  \centering \subfigure[\label{fig:contract_conn}]%
  {\includegraphics[width=0.25\linewidth]{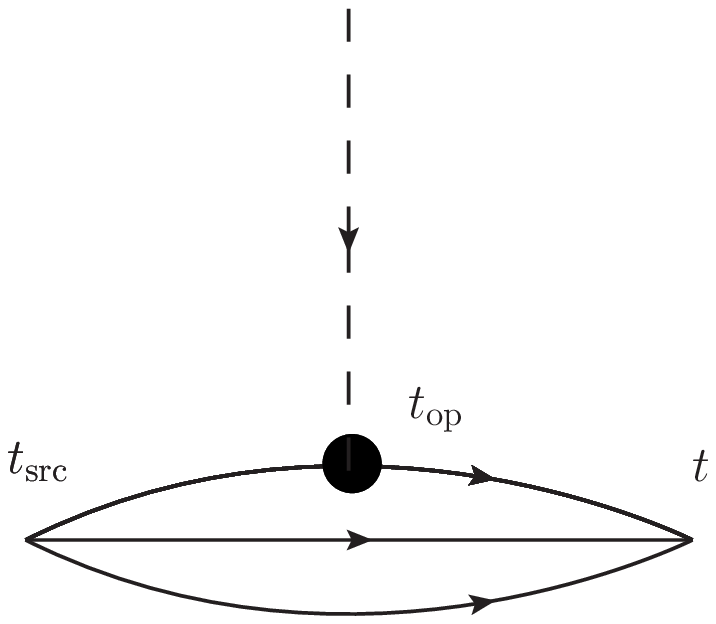}}
  \qquad\qquad \subfigure[\label{fig:contract_disc}]%
  {\includegraphics[width=0.25\linewidth]{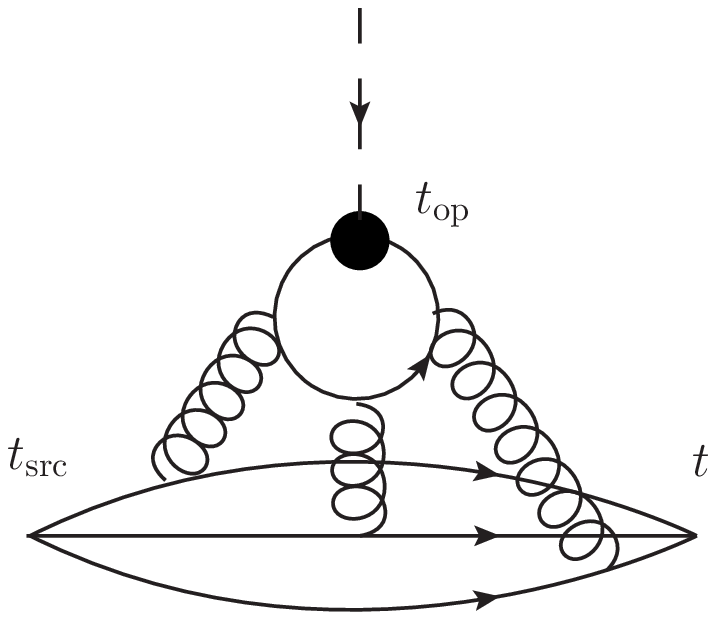}}
\vspace{-0.5cm}  \caption{We illustrate the connected (left) and the disconnected
   (right) graphs that arise from the contractions leading to the 3-point function 
   discussed in the text.}
  \label{fig:contract}
\end{figure}

Since we consider an operator with a non-vanishing vacuum 
expectation value, we also define
\be\label{eq:C3pts_vev_sub}
C^{\pm,O_q,\rm{vev}}_{N,\rm 3pts}(t_{s},\Delta t_{\rm{op}},\vec{x}_{\rm src }) =           
C^{\pm,O_q}_{N,\rm 3pts}(t_{s},\Delta t_{\rm{op}},\vec{x}_{\rm src }) -      
C^{\pm }_{N,\rm 2pts}(t, \vec{x}_{\rm src})   \sum_{\vec{x}_{\rm op}} \la O_q(x_{\rm op}) \ra\; .
\ee
The desired scalar matrix elements can then be computed using the asymptotic 
behaviour of the ratio of a three and two-point functions:
\be\label{eq:ratio_def}
R_{O_q}(t_s,\Delta t_{\rm op}) = \frac{C^{+,O_q,\rm{vev}}_{N,\rm 3pts}(t_s,t_{\rm op})}{C^{+}_{N,\rm 2pts}(t, x_{\rm src}) }  = \la N | \bar{q} q | N \ra + \mcO( e^{-\Delta  t_{\rm op}}) +  \mcO( e^{-\Delta (t_s - t_{\rm op})} )\; .\ee
The general form of the 3-point functions in \eq{eq:C3pts}
lead to both, connected ($\widetilde{C}$, illustrated in fig.~\ref{fig:contract}a) 
and disconnected ($\mcD$, illustrated in \ref{fig:contract}b) contributions, 
\be
C^{\pm,O_q}_{N,\rm 3pts}(t_{s},\Delta t_{\rm{op}}) = 
\widetilde{C}^{\pm,O_q}_{N,\rm 3pts}(t_{s},\Delta t_{\rm{op}}) +  \mcD^{\pm,O_q}_{N}(t_s,t_{\rm{op}},x_{\rm src})
\label{eq:3pt}
\ee
In the following we will denote by $R_{\rm{conn.}}$ (resp.  $R_{\rm{disc.}}$) the 
contribution of $\widetilde{C}^{\pm,O_q}$ (resp. $\mcD^{O_q}$) to the ratio defined 
in \eq{eq:ratio_def}. The sum of the connected and disconnected contribution to the ratio will be denoted $R_{\rm{full}}$.

In order to improve the signal over noise ratio, we have averaged 
the disconnected part over forward and backward propagating
proton and neutron states.    
In addition, we have used up to $4$ randomly chosen source points
per configuration for the 2-point function computation to enhance our statistics.

As will be detailed in a forthcoming publication \cite{paper}, the operators $\widetilde{O}_q$ do not mix under 
renormalization and hence have a straightforward
renormalization pattern very similar to chirally invariant overlap fermions. 
We consider this fact as a major advantage of our twisted mass approach 
for computing the scalar quark contents of the nucleon.  \vspace{-0.25cm}

\newpage

\section{Variance reduction}

The main challenge to compute the 3-point functions of \eq{eq:C3pts} is to evaluate  
$\mcD^{\pm,O_q}$. Our strategy is based on a variance reduction technique  
for twisted mass fermions introduced in ~\cite{Michael:2007vn} and used to  study 
the $\eta'$ meson in ~\cite{Jansen:2008wv}. It relies on the fact that in the 
twisted basis the disconnected contribution that has to be evaluated is 
related to the difference of $1/D_{q,-}- 1/D_{q,+}$.

Here, we will employ the one-end-trick \cite{Michael:2007vn} to compute the disconnected
distribution 
stochastically evaluating 

\be\label{eq:vv_method_disc}
 2i  a\mu_q \sum_{\vec{x}} \left[ \phi_{[r]}^*(x) \gamma_5 \Gamma  \phi_{[r]}(x) \right]_R  =  \sum_{\vec{x}} {\bf{tr}}~ \Gamma \left( \frac{1}{D_{q,-}}- \frac{1}{D_{q,+}}\right)(x,x) + \mcO\left( R^{-1/2}\right),
\ee
where  
\be
 \phi_{[r]}=(1/D_{q,+}) \xi_{[r]} \gap\textmd{and}\gap \phi_{[r]}^* = \xi_{[r]}^* (1/D_{q,+})^{\dag}\; ,
\ee 
where we have introduce $N_R$ independent random volume sources  denoted 
$\xi_{[r]}$. For the generation  of the random sources we have used 
a $\mathbf{Z}_2$ noise setting all field components randomly from the set $\{1,-1\}$.  

\begin{figure}[h]
\begin{minipage}[ht]{7.5cm}
\includegraphics[height=6cm,width=7.5cm]{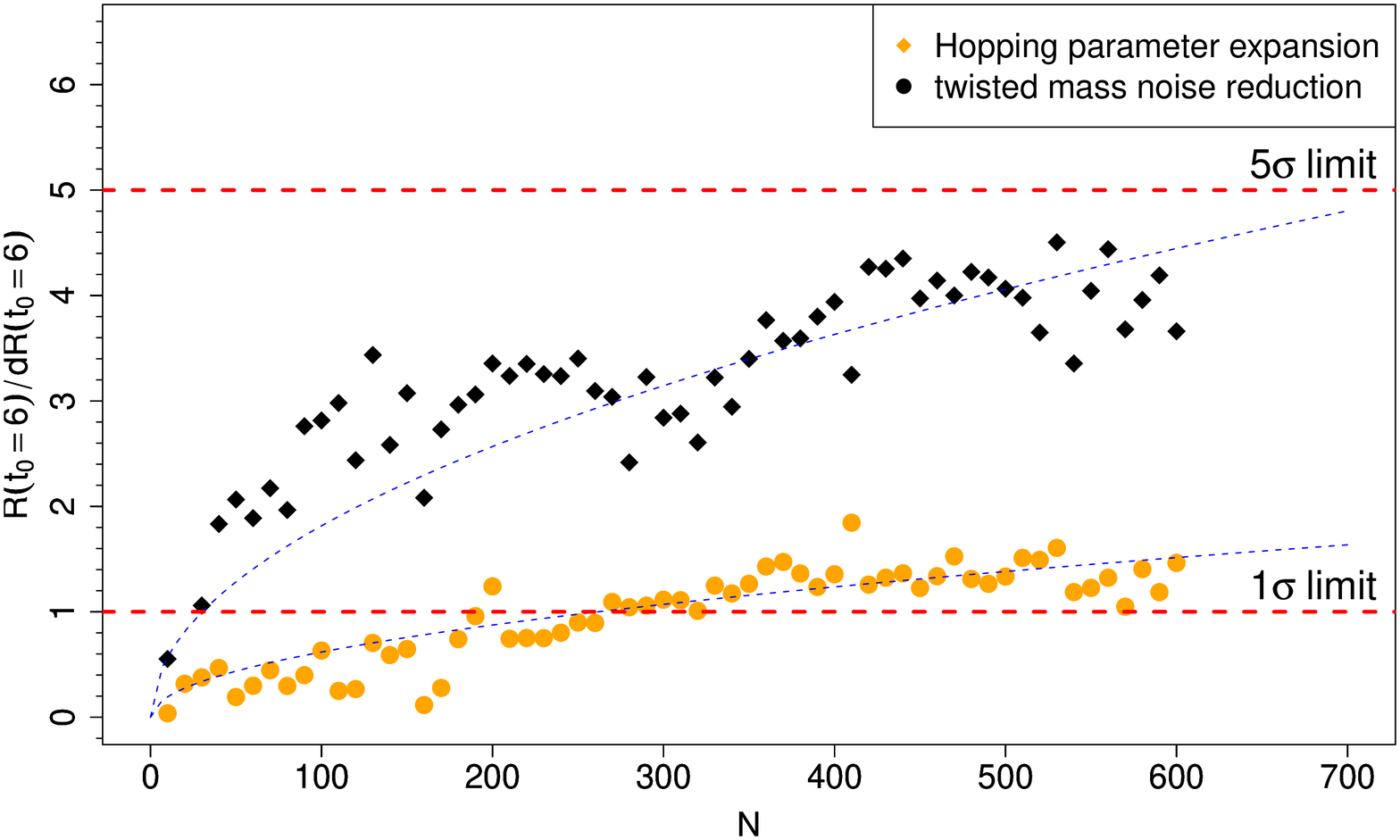}
\label{fig:rel_err_scaling}
\vspace*{-1cm}
\caption{Signal to noise ration (SNR) of the ratio $R_{O_q}$ 
for a fixed time $\Delta t_{\rm op}=t_s/2 = 6$ as a function of the number of gauge field configurations $N$. 
$R_{O_q}$ is evaluated here  in the strange quark regime.
We compare our noise reduction technique, with the hopping parameter expansion technique.
The dashed lines indicate the $1\sigma$ and $5\sigma$ significance levels and 
the short dotted lines are only shown to guide the eyes.}
\end{minipage}
\hspace{0.5cm}
\begin{minipage}[ht]{7.5cm}\vspace*{-1.7cm}
\includegraphics[height=6cm,width=7.5cm]{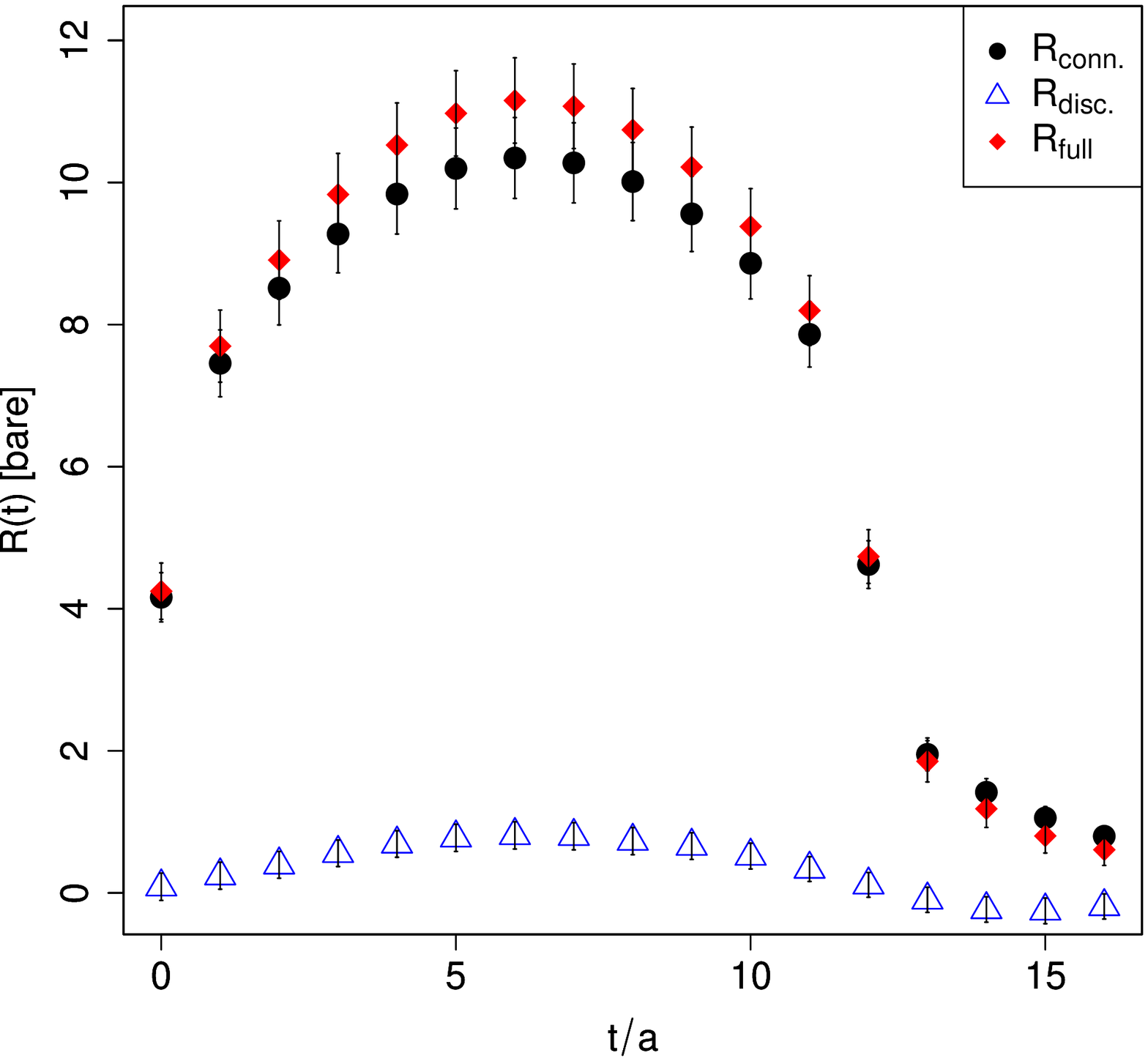}
\caption{Plot of the two contributions $R_{\rm{disc.}}$ and $R_{\rm{conn.}}$  to the ratio 
$R_{O_q}$ relevant for the extraction of $\sigma_{\pi N}$. $R_{\rm{full}}$ is 
the sum of both contributions. 
}\label{fig:_bare_light_plateau}
\end{minipage}
\end{figure}

In our tests for the signal to noise ratio (SNR) we first investigated 
how the SNR depends on the number of stochastic  
sources $N_R$ used. We found that for $N_R\gtrsim 7$ 
there is no significant improvement of the SNR. 
Nevertheless, we have
used $12$ stochastic sources per configurations in all our results.  
In \fig{fig:rel_err_scaling} we compare the SNR of the twisted mass specific variance 
reduction technique to a more standard method based on the the hopping parameter 
expansion\cite{McNeile:2000xx}. We show the ratio $R_{O_q}$ of eq.~(\ref{eq:ratio_def}), for a fixed value 
of $\Delta t_{\rm op}=t_s/2 = 6$  as a function of the number of configurations $N$. 
We conclude that the SNR is increased by a factor $\sim 3$ with our improved noise 
reduction technique which allows to obtain a result at the $5\sigma$ significance 
level with only a moderate statistics. 
\vspace*{-0.7cm}

\vspace*{-0.6cm}
\section{Light $\sigma-$term}

In \fig{fig:_bare_light_plateau} we show the results obtained for the bare ratio $R_{\rm{full}}$  
introduced in \eq{eq:ratio_def} from which $\sigma_{\pi N}$ can be computed. 
The connected part, $R_{\rm{conn.}}$, for a source-sink separation of $t_s=12a$,  is shown by 
the black filled circles. It has been computed using the standard method of ``sequential inversions through the 
sink''. As can be seen,  $R_{\rm{conn.}}$ shows a 
time dependence indicating excited state contributions, a systematic effect 
that has not been taken into account yet  
in this work. The disconnected part, $R_{\rm{disc.}}$, is represented by 
blue triangles in \fig{fig:_bare_light_plateau}. The disconnected part is significantly 
smaller than the connected part $R_{\rm{conn.}}$ and contributes at the 
$\sim 10\%$ level to the full ratio $R_{\rm{full}}$ represented by the red diamonds.  

We have computed  $\sigma_{\pi N}$ so far for three values of the pion mass, 
see table~\ref{tab:sigma_piN}. Our present data do not allow for a reliable extrapolation to the physical
point for which additional data at more pion masses would be necessary.  
\vspace*{-0.5cm}
\begin{table}[htb]
\begin{center}
\begin{tabular}{|c|c|}
  \hline
  $\mps$ ($\mev$) & $\sigma_{\pi N}$ ($\mev$) \\
  \hline
  318 & 99(6) \\
 392 & 152(9) \\
 455 & 228(15) \\
 \hline   
\end{tabular}
\caption{Fit results for   $\sigma_{\pi N}$ as a function of the pion mass. Only statistical error are estimated.}
\label{tab:sigma_piN}
\end{center}
\end{table}
%


%

\vspace*{-0.5cm}\section{Strangeness of the nucleon}

\begin{figure}[h]
\vspace*{-1.25cm}\begin{minipage}[ht]{7.cm}
\includegraphics[height=6cm,width=6cm]{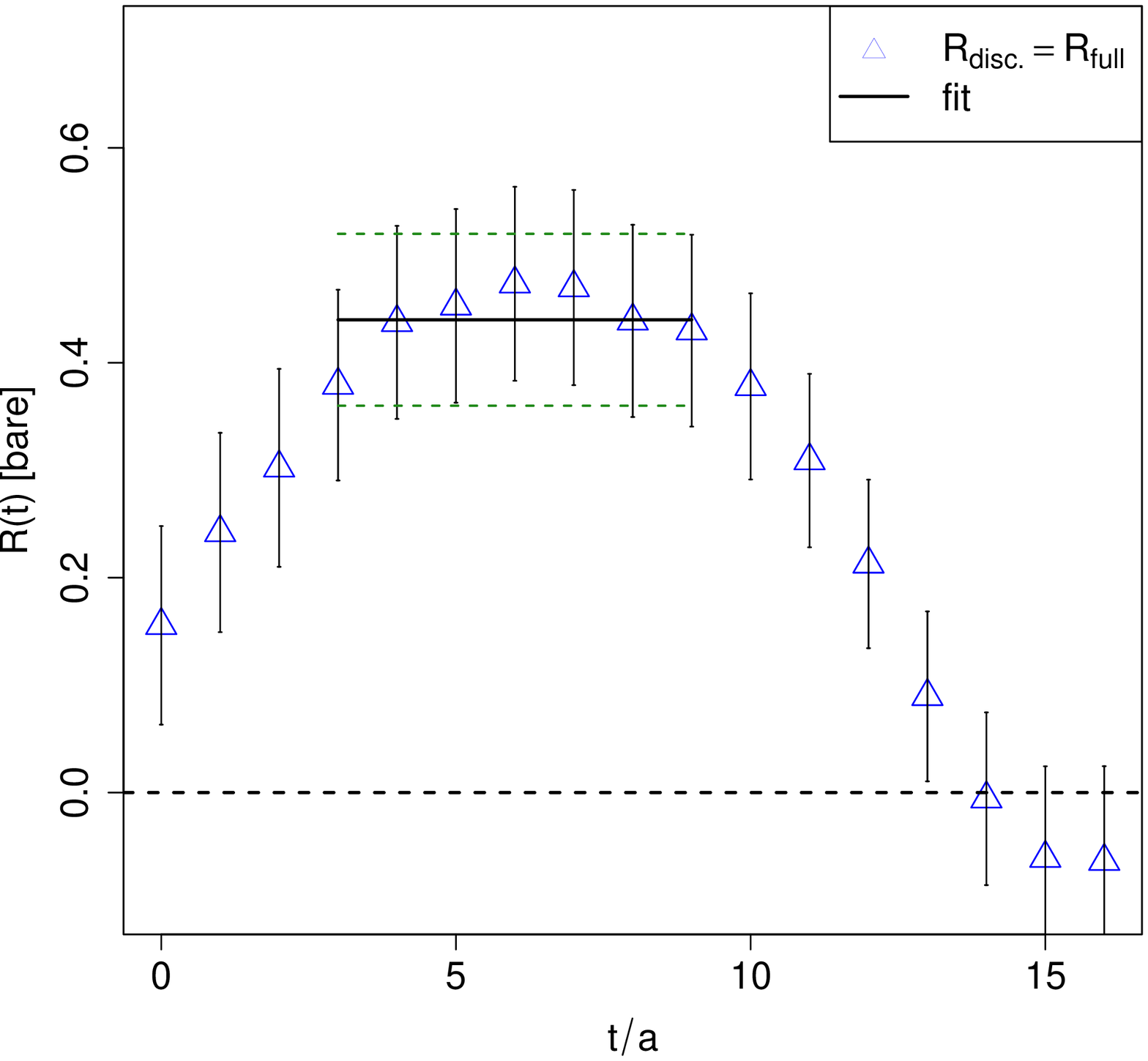}
\vspace*{-0.35cm}\caption{Time dependence of $R_{\rm{disc.}}$ in the strange quark mass regime 
($a\mu_q=0.018$). The source-sink separation has been fixed to $12a$ ($842$ configurations)}\label{fig:plateau_strange}
\end{minipage}
\hspace{0.5cm}
\begin{minipage}[ht]{7.cm}\vspace*{-0.5cm}
\includegraphics[height=6cm,width=6.cm]{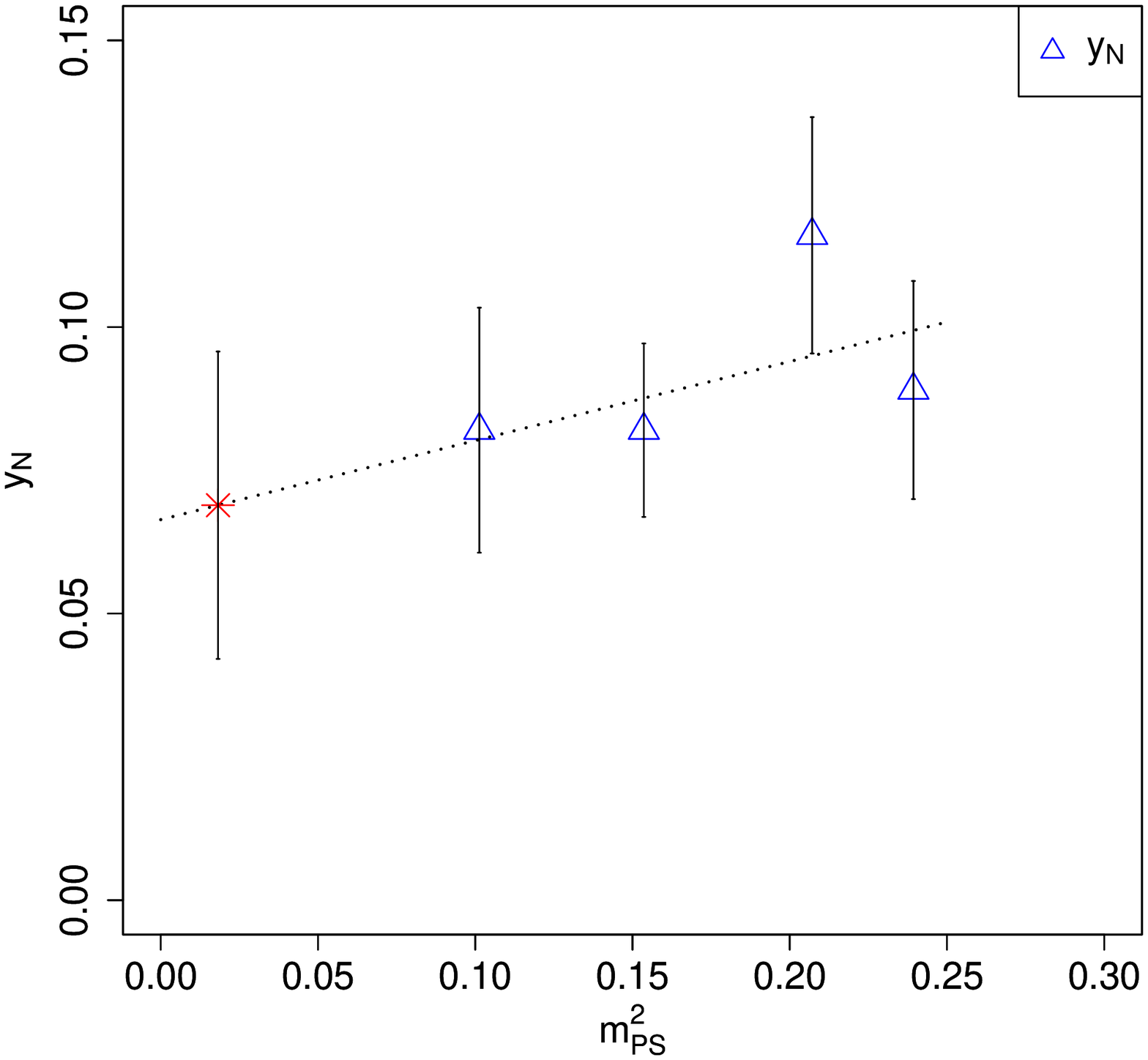}
\vspace*{-0.5cm}\caption{Our data for $y_N$ as a function of the pion mass.  
The data are extrapolated to the chiral limit using a simple 
linear extrapolation. }\label{fig:xfit_y}
\end{minipage}
\end{figure}

In \fig{fig:plateau_strange}, we show $R_{\rm{disc.}}$ for a quark 
mass of $a\mu_q = 0.018$ corresponding approximately to the strange quark mass.  
The in principle freely selectable source-sink separation has been fixed to 
$12$ lattice units. 
The ratio $R_{O_q}$ of eq.~(\ref{eq:ratio_def}) shows a time dependence 
indicating that also in the case of the strange quark excited 
states may be important. We nevertheless extract a plateau value as
indicated in \fig{fig:plateau_strange} which is clearly different from 
zero.  
Combining this value with the result for the scalar matrix 
element obtained in the light quark sector discussed above, 
allows us finally to compute $y_N$.   
We have performed such a computation at four values of the pion mass 
at fixed value of the lattice spacing and the results are summarized in \fig{fig:xfit_y}. 
Performing a simple  linear extrapolation we obtain, as shown by a red star, $y_N = 0.069(27)$ in the chiral limit, where only the statistical 
errors have been taken into account. 
Although an investigation of systematic effects are still 
missing, we checked that by varying the bare strange quark mass to $a\mu_s = 0.016$ does not 
change significantly the value of $y_N$ . 

As a final remark we mention that we have also computed the charm quark content
of the nucleon. 
Unfortunately, here the SNR is of order one and hence no clear signal 
can be extracted. From our present data we can only provide 
a qualitative estimate that $\la N | O_c| N \ra \lesssim \la N | O_s | N \ra$. 

\vspace*{-0.5cm}
\section{Conclusion}
\vspace*{-0.5cm}
In this proceedings contribution we have shown that with twisted mass fermions
it is possible address the disconnected contribution to the scalar 
matrix elements of the nucleon. This becomes especially important 
when the strange and the charm content are computed since there 
only disconnected graphs appear.  
In addition, with twisted mass fermions the renormalization pattern  
of such matrix elements 
is the same as for chirally invariant discretizations. As a result 
we were able to obtain accurate  
results in the light and strange sector at fixed lattice spacing and 
for several quark masses. Our main result is a value 
$y_N =  0.069(27)$. This value is  compatible with recent lattice results 
obtained by several groups\cite{JLQCD,BMW,QCDSF}. The still missing systematic uncertainties 
on this result will be addressed in future simulations. 

\vspace*{-0.5cm}
\section*{Acknowledgments}
This work was performed using HPC resources provided by the JSC Forschungszentrum J\"ulich on the JuGene supercomputer and by GENCI (IDRIS-CINES) Grant 2010-052. It is supported in part by  the DFG Sonder\-for\-schungs\-be\-reich/ Trans\-regio SFB/TR9.

\end{document}